\documentclass[twocolumn,superscriptaddress,amsmath,amssymb,floatfix,prl]{revtex4}

\usepackage{graphicx}

\def\nm{{\ {\rm nm}}}						
\def\micron{{\ \mu{\rm m}}}					

\def\kHz{{\ {\rm kHz}}}						
\def\MHz{{\ {\rm MHz}}}						

\def\us{{\ \mu{\rm s}}}						
\def\ms{{\ {\rm ms}}}						

\def\Er{{{E_R}}}							
\def\kr{{{k_R}}}							
\def\Rb87{^{87}\rm{Rb}}					
\def\ToverUC{(t/U)_{\rm c}}					

\def\nbar{\left<{\hat n}_k\right>}					
\def\nbarexpt{\left<n(k_x,k_y)\right>}				

\newcommand{\ket}[1]{|#1\rangle}

\begin{document}

\title{The Mott insulator transition in a two dimensional atomic Bose gas}

\author{I.~B.~Spielman}
\email{ian.spielman@nist.gov}
\affiliation{National Institute of Standards and Technology, 
Gaithersburg, Maryland, 20899, USA}
\author{W.~D.~Phillips}
\affiliation{National Institute of Standards and Technology, 
Gaithersburg, Maryland, 20899, USA}
\affiliation{University of Maryland, College Park, Maryland, 20742, 
USA}
\author{J.~V.~Porto}
\affiliation{National Institute of Standards and Technology, 
Gaithersburg, Maryland, 20899, USA}
\affiliation{University of Maryland, College Park, Maryland, 20742, 
USA}

\date{\today}

\begin{abstract}
Cold atoms in periodic potentials are remarkably versatile quantum systems for implementing simple models prevalent in condensed matter theory.  Here we realize the 2D Bose-Hubbard model by loading a Bose-Einstein condensate into an optical lattice, and we study the resulting Mott insulating state (a phase of matter in which atoms are localized on specific lattice sites).  We measure momentum distributions which agree quantitatively with theory (no adjustable parameters).  In these systems, the Mott insulator forms in a spatially discrete shell structure which we probe by focusing on correlations in atom shot noise.  These correlations show a marked dependence on the lattice depth, consistent with the changing size of the insulating shell expected from simple arguments.
\end{abstract}

\maketitle

In recent years, ultra-cold atoms confined in optical lattices have realized fascinating strongly interacting condensed matter phenomena, including the Girardeau-Tonks gas in 1D~\cite{Paredes2004,Kinoshita2004} and the superfluid to Mott insulator transition in 3D, and 1D~\cite{Greiner2002,Stoferle2004}.  The Mott-insulating state is a strongly correlated phase of matter where interactions localize the constituent particles to individual lattice sites.  While the existence of a 2D Mott insulator has been verified in the cold atom system~\cite{Kohl2005}, it has gone largely unexplored.  In all of these systems, an atomic Bose-Einstein condensate (BEC) is loaded into the optical lattice formed by interfering laser beams, confining the atoms in a nearly perfect 1, 2, or 3D periodic potential.  The atom-optical system differs substantially from traditional condensed matter systems both in control (the trapping potentials can be changed dynamically on all relevant experimental time scales) and in measurement opportunities (e.g., imaging after time-of-flight provides a direct measurement of the momentum distribution).  Imaging complements techniques usually available in condensed matter systems, and allows us to study the 2D Mott insulator via its momentum distribution and correlations in its noise~\cite{Altman2004,Folling2005,Greiner2005,Schellekens2005}.  Our measured momentum distributions agree quantitatively with the predictions of a 2D theory with no adjustable parameters (even approaching the insulator-superfluid transition, where perturbation theory breaks down).  In the cold atom system, the transition from superfluid to insulator occurs smoothly; the system segregates into shell-like domain(s) of insulator and superfluid~\cite{Jaksch1998,Batrouni2002,Folling2006,Campbell2006}.  Even in the absence of direct imaging, we find signatures of the insulating shell in noise-correlations.  These noise-correlations dependence on the lattice depth, trending as expected due to the changing size of the Mott insulator region.

Our sample consists of an ensemble of 2D Bose systems in a combined sinusoidal plus harmonic potential.  Absent the harmonic confinement, the system is well described by the homogeneous Bose-Hubbard Hamiltonian~\cite{Fisher1989,Jaksch1998}, which models bosons on a lattice in terms of $t$, the matrix element for tunneling between sites, and $U$, the on-site energy cost for double occupancy.  At sufficiently low temperature, this system is either a superfluid (SF) or a Mott insulator (MI).  In 2D, the SF state has quasi-long range phase coherence and a gapless excitation spectrum linear in momentum; in contrast, the MI phase has a gap between the ground and first excited states.  For unit filling (on average one atom per lattice site), the 2D Bose-Hubbard system exhibits a zero temperature quantum phase transition from SF to MI~\cite{Fisher1989} when the dimensionless ratio $t/U \approx 0.06$~\cite{Elstner1999,Wessel2004}, which we denote by $\ToverUC$.  When $t/U$ vanishes, the unit filled MI has exactly one atom per-lattice site.  At finite $t/U$ this is not fixed and the variations from unit occupation are correlated between nearby sites.  This is associated with coherence between these sites, resulting in broad, diffractive, structure in the momentum distribution~\cite{Gerbier2005}.

\begin{figure*}[t!]
\begin{center}
\includegraphics[width=6.75in]{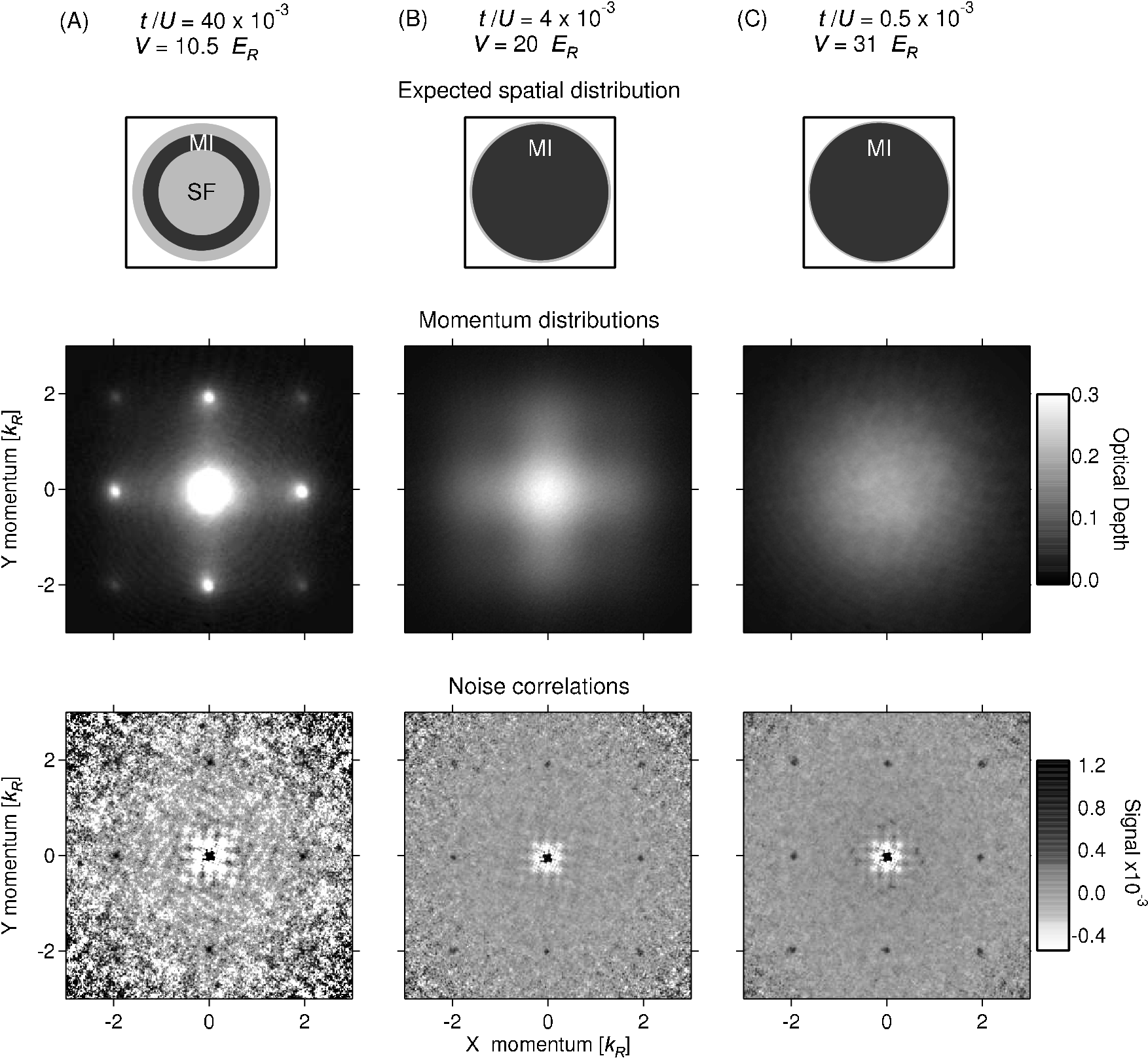}
\end{center}
\caption{Results for 3 different values of $t/U$.  Top row: expected in situ density profile from a LDA calculation using the 2D MI phase diagram from ref.~\cite{Elstner1999}.  The regions of dark grey denote the MI phase, and the light regions indicate SF.  In all cases, the MI is surrounded by a small ring of SF.  Middle row: imaged atom density versus momentum.  Bottom row: noise correlations of the images as a function of momentum difference.  Each displayed image was averaged from about 60 raw images; also, to reduce the noise, the noise-correlation data shown was averaged with itself, rotated by $90^\circ$.  While the momentum distributions at $t/U = 4\times10^{-3}$ and $0.5\times10^{-3}$ are quite different, the shell structures hardly differ.}
\label{rawdata}
\end{figure*}

In current experiments, the additional harmonic trapping potential modifies the physics from the ideal case described by the homogeneous Bose-Hubbard Hamiltonian.  In the trapped system, domains of the insulating phase grow continuously as $t/U$ decreases (we vary $t/U$ by changing the depth of the lattice potential), forming a spatially discrete shell structure~\cite{Jaksch1998,Batrouni2002,Folling2006,Campbell2006}.  This is most simply interpreted in terms of a ``local chemical potential'' which varies from a peak value in the center of the trap to zero at the edges (i.e., a local density approximation, LDA).  In this case, the system segregates into one or more shells of MI separated by SF.  In our experiment, the coexisting phases are SF and unit-occupancy MI.  Examples relevant to our experiment are shown schematically in the top row of Fig.~\ref{rawdata}, where (A) is a system which has a SF core enveloped by a MI shell.  (B) and (C) represent deeper lattices where the sample is MI nearly throughout.

We prepare the Bose-Hubbard system at a specific value of $t/U$~\footnote{We compute $U$ and $t$ in a 2D band structure calculation using the measured lattice depth.}, measure the momentum distribution, and extract correlations in its noise.  The characteristic loss of diffraction in the momentum distribution~\cite{Greiner2002} as the system goes deeper into the Mott regime is shown in middle row of Fig.~\ref{rawdata}.  At low temperature, the MI is expected to form in a shell structure, with each region contributing to the image; it can be difficult to deconvolve the separate contributions, particularly near the transition~\cite{Gerbier2005,Gerbier2005a}.  Despite these limitations, we find that a theory for the homogeneous system is surprisingly good at describing the full momentum distribution.

The bottom row of Fig.~\ref{rawdata} shows noise correlations of the momentum distribution, whose diffractive nature does not vanish, even deep in the Mott regime~\cite{Altman2004,Folling2005}.  The width and area of the diffractive noise correlation peaks should depend on the size of the Mott region.  Deep in the Mott limit, the width of the correlation peaks should be diffraction limited by the inverse linear size of the Mott domain, and the area of the correlation peaks should depend on the number of atoms in the MI.  Indeed, we observe that the area of the correlation peaks clearly increases with increasing $t/U$, while the width shows a more subtle trend.

\section{Experimental procedure}

We produce nearly pure 3D $\Rb87$ BECs with $N_T=1.7(5)\times10^5$~\footnote{All uncertainties herein reflect the uncorrelated combination of single-sigma statistical and systematic uncertainties.} atoms in the $\left|F=1,m_F=-1\right>$ state~\cite{Spielman2006}.  The 3D BEC is separated in $200\ms$ into an array of about $60$ 2D systems by an optical lattice aligned along $\hat z$ (vertical lattice).  This lattice is formed by a pair of linearly polarized $\lambda = 820\nm$ laser beams~\footnote{We have replicated these experiments with our laser tuned to $810\nm$ with the same results.  Our results for the noise correlation measurements include both data sets.}.  In addition, a square 2D lattice in the $\hat x$-$\hat y$ plane is produced by a second beam arranged in a folded-retroreflected configuration~\cite{Sebby-Strabley2006}; the polarization is aligned in the $\hat x$-$\hat y$ plane.  This $\hat x$-$\hat y$ lattice is gradually applied in $100\ms$~\footnote{The vertical lattice beams intersect at $\theta=162(1)^\circ$ giving a lattice period of $415(1)\nm$, compared to the $410(1)\nm$ periodicity of the $\hat x$-$\hat y$ lattice.  The beams for these two lattices originate from the same laser but differ in frequency by about $80\MHz$.}.  The intensities of the vertical and $x$-$y$ lattices follow exponentially increasing ramps (with $50\ms$ and $25\ms$ time constants respectively) which reach their peak values concurrently.  These time-scales are chosen to be adiabatic with respect to mean-field interactions, vibrational excitations, and tunneling within each 2D system \footnote{When the lattice depths exceeds $\sim 25 \Er$, the loading ceases to be adiabatic with respect to tunneling~\cite{Gericke2006}.  The vertical lattice is always in this limit, leading to a final ensemble of 2D systems whose respective phases are randomized.}.  Ideally, the loading process transfers a ground state BEC to the ground state of the lattice system.  (While we do not know the temperature in the lattice, we begin with a nearly pure BEC and are confident that our loading procedure does not cause excessive heating.  To check this adiabaticity, we created a MI as described below, then decreased the lattice potential in about $30\ms$ and verified the reappearance of the sharp diffraction orders indicative of the SF phase.)  The final vertical lattice depth is always $30(2)\Er$; the single photon recoil momentum and energy are defined as $\kr = 2\pi/\lambda$ and $\Er=\hbar^2\kr^2/2 m=h\times3.4\kHz$.  The final depth of the $\hat x$-$\hat y$ lattice determines $t/U$ and ranges from $V=0$ to $31(2)\Er$.  In our experiments, the lattice depths are measured by pulsing the lattice for $3\us$ and observing the resulting atom diffraction~\cite{Ovchinnikov1998}.

Once both lattices are at their final intensity, the atomic system consists of a set of 2D gasses each in a square lattice of depth $V$ and with a typical density of 1 atom per lattice site.  The atoms are held for $30\ms$ at which point all confining potentials are abruptly removed (the optical lattice and magnetic potentials turn off in $\lesssim1\us$ and $\simeq300\us$, respectively).  When the potentials are thus ``snapped off'' the initially confined states are projected onto free particle states which expand for a $20$ to $30\ms$ time-of-flight (TOF); they are then detected by resonant absorption imaging (see the supporting Methods section for details).  Because initial momentum maps into final position, each image corresponds to a measurement of the 2D momentum distribution $n(k_x,k_y)$.  Ideally, when averaged over many realizations, this converges to the expectation value $\left<{\hat n}_{k_x,k_y}\right>$ of the 2D density operator.

The middle row of Fig.~\ref{rawdata} shows a series of such averages, at different lattice depths, starting near the MI transition and crossing deep into the MI phase~\cite{Greiner2002}.  These images illustrate the smooth progression from sharp diffraction peaks on a small background (A), to broad diffraction peaks with a considerable background (B), culminating with a near-perfect Gaussian distribution (C).  Because the system is divided into domains of SF and MI, the diffraction images can be difficult to interpret -- only returning to simplicity in the limits of a shallow or deep lattice when the entire system consists of just one phase (SF or MI).  Understanding quantitatively the intermediate regime requires careful modeling of the inhomogeneous density distribution~\cite{Gerbier2005a}.

\begin{figure}[t!]
\begin{center}
\includegraphics[width=3.375in]{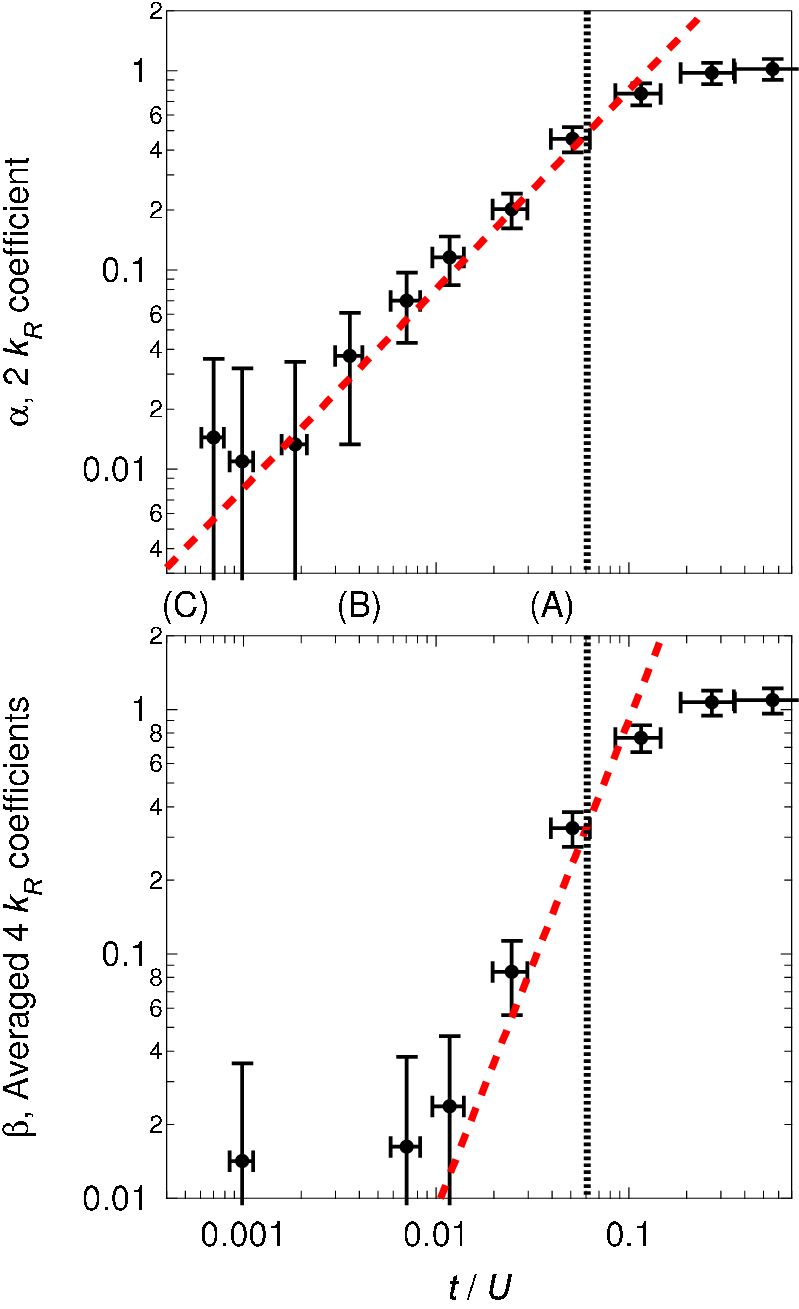}
\end{center}
\caption{Top: the symbols show the measured first order coefficient $\alpha$ versus $t/U$.  For reference the red dashed line shows the expected value $\alpha = 8 t/U$.  Each point is acquired from a single image at known lattice depth, therefore known $t/U$.  (A), (B), and (C) indicate the value of $t/U$ for the three sets of data in Fig.~\ref{rawdata}.  Bottom: the symbols denote the averaged second order coefficient $\beta$ versus $t/U$.  The red dashed line is the expected power-law $\beta=90(t/U)^2$.  In both figures the vertical dotted line at $\ToverUC=0.06$~\cite{Elstner1999,Wessel2004}, indicates the expected location of the 2D SF-MI transition.  (Some extracted values of $\beta$ are negative, and to not appear on the log plot.)
}
\label{ToverU}
\end{figure} 

\section{Momentum distributions}

In the deep lattice limit, the MI wave function $\ket{\Psi_0}$ has an exact number of atoms in each occupied site (a Fock state) -- in our case $n=1$.  At large but finite depth, the unit-occupied state is modified to first order in perturbation theory with a small mixture of neighboring particle-hole pairs.  As in a 3D lattice, this gives a modulated momentum distribution, here $\nbar = N \left|w(\bf k)\right|^2 \left\{1 + \alpha\left[\cos(\pi k_x/k_r) + \cos(\pi k_y/k_r)\right]  \right\}$~\cite{Gerbier2005,Gerbier2005a}.  Here $\alpha = 8 t/U$, and $w({\bf k})$ is the Fourier transform of the localized Wannier states in the optical lattice sites; $w({\bf k})$, determined by the lattice depth, is well approximated by a Gaussian for all data described here.  Because the density modulations result from an interference between the unit-occupied Mott state and the particle-hole admixture, the lowest order correction to $\nbar$ is proportional to $t/U$, even though the probability for double occupancy scales as $(t/U)^2$.

This modulation in density has been verified to be first order in $t/U$ over a range of parameters in 3D~\cite{Gerbier2005}.  As with the 2D case, which we discuss below, this agreement near the MI-SF transition can be surprising for two reasons: (1) as seen in the top row of Fig.~\ref{rawdata}, only a fraction of the inhomogeneous system may be in the MI phase; (2) as $t/U$ approaches the critical value from below, higher order contributions become important. 

To quantify the next order term, we expand an analytic result (in the random phase approximation, Ref. ~\cite{Sengupta2005}) to second order in $t/U$.  The expected second order term in the momentum distribution is $72 (t/U)^2 [\cos(\pi k_x/k_r) + \cos(\pi k_y/k_r)]^2$.  This yields additional Fourier terms $\beta_1(\cos(2\pi k_x/k_r)+\cos(2\pi k_y/k_r))$, and $\beta_2\cos(\pi k_x/k_r)\cos(\pi k_x/k_r)$; we define the average coefficient $\beta = (\beta_1+\beta_2)/2 = 90 (t/U)^2$.  (Terms higher order in $t/U$ also contribute slightly to these Fourier components.)

We extract the Fourier coefficients $\alpha$ and $\beta$ from individual images and plot them in Fig.~\ref{ToverU}.  The top panel shows $\alpha$ versus $t/U$; the red dashed line is the theoretical prediction with no adjustable parameters: $\alpha=8 t/U$.  The vertical black line denotes $\ToverUC=0.06$, where we expect the MI to first appear somewhere in our sample.  Points to the left of the black dotted line (MI regime) agree with perturbation theory.  

The bottom panel of Fig.~\ref{ToverU} shows the $\beta$ versus $t/U$.  The red dashed line is the predicted value $\beta=90(t/U)^2$.  In the MI regime, our measurements agree with expectations, within experimental uncertainty.  Note that near the MI transition the first and second order term become comparable, indicating the incipient breakdown of perturbation theory.

\begin{figure}[t!]
\begin{center}
\includegraphics[width=3.375in]{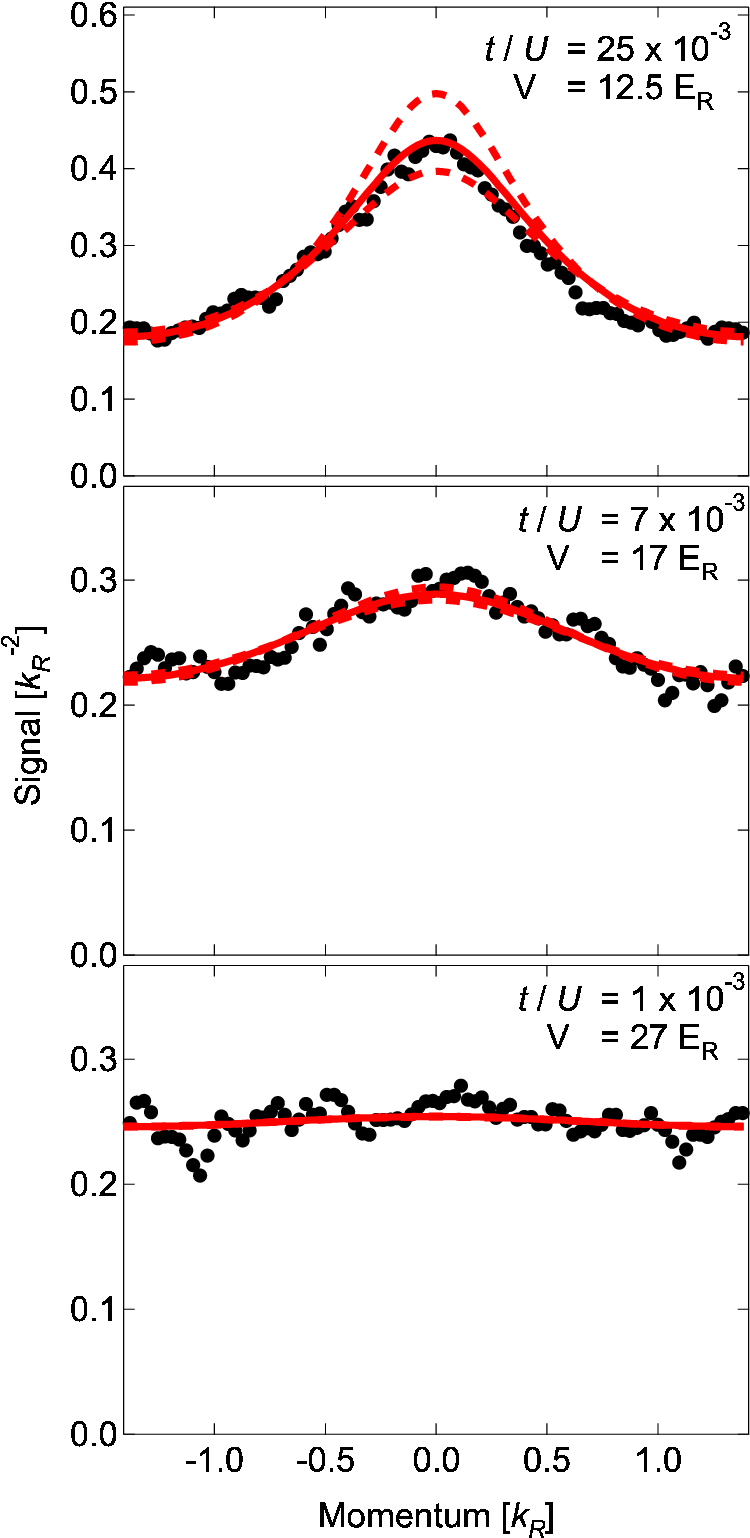}
\end{center}
\caption{Cross sections of normalized quasi-momentum distributions (along $\hat x + \hat y$) at three values of $t/U$ (along this diagonal the section extends to $k=\pm\sqrt{2}$).  In each case the data is plotted along with the theoretical profile (red lines)~\cite{Sengupta2005}.  The dashed lines (not visible in the bottom panel) reflect the uncertainty in the theory resulting from the single-shot $\pm0.5\Er$ uncertainty in the lattice depth.
}
\label{ModelSections}
\end{figure} 

In Fig.~\ref{ModelSections}, we directly compare the measured quasi-momentum distribution with theory for three values of $t/U$.  The modulated momentum distribution is the product of the  magnitude squared Wannier function $|\omega(k)|^2$ and the quasi-momentum distribution (periodic along $\hat x$ and $\hat y$ in the reciprocal lattice vectors, $2\kr$).  To extract the normalized quasi-momentum distribution we divide the data by $|\omega(k)|^2$, then calculate a properly weighted average of all points separated by multiples of the reciprocal lattice vectors, and finally normalize.  The data in Fig.~\ref{ModelSections} are cross sections of the quasi-momentum distribution along $\hat x+\hat y$ plotted with the prediction of the RPA theory (red lines).  At low $t/U$ the quasi-momentum distributions are cosinusoidal.  As $t/U$ increases toward $\ToverUC$, contributions of higher Fourier terms become important, as is evident in the cross section at $t/U=25\times10^{-3}$ which deviates from a cosine.  The shape of the measured distribution matches the predictions of theory with no free parameters.

\section{Noise correlations}

Although our sample is inhomogeneous, a homogeneous-system theory provided a remarkably good representation of the data discussed above.  This results from two facts: (1) except quite close to the MI transition, nearly all of the system should be unit occupied MI, and (2) by focusing only on the largest momentum scales (corresponding to spatial length scales on the order of one- or two- lattice sites) we are insensitive to the size of the MI.  In contrast, the size of the MI almost exclusively determines the area and width of the peaks in the noise correlations signal (bottom row of Fig.~\ref{rawdata})~\cite{Altman2004,Folling2005}.

We determine noise correlations from our images of atom density $n(k_x,k_y)$ by computing the autocorrelation function averaged over many images: 
\begin{equation}\nonumber
S(\delta k_x, \delta k_y) = 
S_0^{-1}\left<\int n(k_x,k_y) n(k_x+\delta k_x, k_y + \delta k_y)dk_x dk_y\right>; 
\end{equation}
we normalize by the autocorrelation of the average $\left<n(k_x,k_y)\right>$, so $S_0(\delta k_x, \delta k_y)=\int\left<n(k_x,k_y)\right>\left<n(k_x+\delta k_x, k_y + \delta k_y)\right> dk_x dk_y$.  In other words, we determine $S(\delta k_x, \delta k_y)$ by calculating the autocorrelation function (ACF) of each image separately, and then average over many realizations, typically (40 to 80).  (We normalize this by the ACF of the averaged images to remove the dependence of $S(\delta k_x, \delta k_y)$ on the momentum distribution $\left<n(k_x,k_y)\right>$.)  It can be shown that this quantity has diffractive structure when $\ket{\Psi_0}$ is a pure Fock state~\cite{Altman2004,Rey2005}, with noise correlation peaks separated by $2\kr$ (the reciprocal lattice vector).  The spacing of the peaks therefore reveals the underlying lattice structure, even in the absence of diffraction in the momentum distribution.  In addition, the correlation peak areas $A$ and widths $\delta k$, provide information about the system.

In the limit of a deep lattice, the ground state of our system is a 3D array of lattice sites with exactly one atom per site.  As is usual for diffraction phenomena, the width of the correlation peaks is determined by the size of the array, and is proportional to $L^{-1}$; where $L\propto N^{1/3}$ is the linear extent of the MI region and $N$ is the number of sources (lattice sites).  Likewise, the area under a peak is related to the atom number in the MI by $A=(2 \kr)^2/N$~\cite{Folling2005}.  As with most noise, the noise-correlation variance (the square of the standard deviation), scales as $N$.  We normalize by a quantity which scales like $1/N^2$, so the relative fluctuations described by $A$ have an overall $1/N$ dependence.  We calculate that this remains true including the order $t/U$ correction to the MI state, so even to first order, noise correlations contain information only about the size of the Mott-insulating system.  (Order $(t/U)^2$ corrections to the noise-correlations, which we have not investigated, may alter this behavior).

\begin{figure}[h!]
\begin{center}
\includegraphics[width=3.375in]{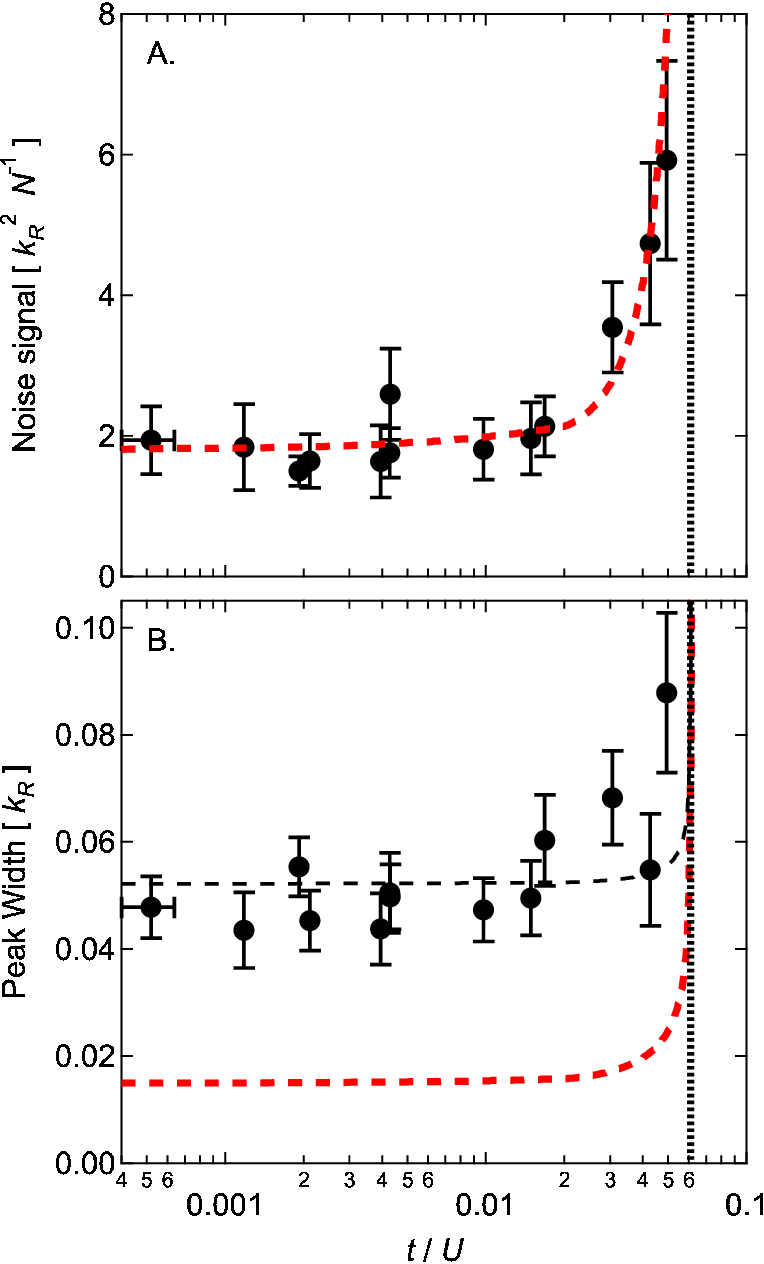}
\end{center}
\caption{Measured average area and width of the two central independent noise-correlation peaks.  The uncertainty bars reflect the statistical uncertainty of the fit due to background noise.  Top: average area expressed in units of $\kr^2 / N_T$.  In these units, the expected area for a perfect $n=1$ Fock state is $4$; the data tend to about $1.8\ \kr^2 / N_T$ in a deep lattice.  The red dashed line denotes the expected area due to the finite number of atoms in the MI as calculated in our LDA model, which was scaled by $0.45$ to lie upon the data.  Bottom: the solid symbols are the measured width in $\kr$ of the correlation peaks.  The red dashed line is the expected peak-width based on our LDA computation of the system size.  The black dashed line shows the expected peak-width including the imaging resolution of $0.05\kr$ (see supporting Methods section).  In both panels, the vertical dotted line show $\ToverUC=0.06$~\cite{Elstner1999,Wessel2004}, the expected location of the 2D SF-MI transition.  
}
\label{NoiseArea}
\end{figure}

As the system approaches the critical point from the MI side, we expect the size of the MI region to shrink, and correspondingly the correlation peaks to become wider, with increasing area (the total number of atoms in the experiment $N_T$ remains fixed; only the fraction of atoms in the MI $N/N_T$ decreases).  Figure~\ref{NoiseArea} shows this general behavior.  Our study of the noise correlations near the MI critical point is enabled by a masking procedure: in each image we eliminate regions of radius $35\micron\approx0.3\times \kr$ centered on the diffraction peaks before computing correlation functions.  This removes spurious effects of the sharp diffraction peaks evident near the MI-SF transition (see supporting Methods section).

Figure~\ref{NoiseArea}A shows the area under the noise correlation peaks measured by fitting the peaks to 2D Gaussians.  The data show that as $t/U$ increases the area also increases; this indicates that the fraction of the system in the MI state is decreasing as expected.  The area expected for small $t/U$ is $(2 \kr)^2/N_T$, however, our data tends to about $45\%$ of this in a deep lattice (a similar suppression of the noise signal was observed in Ref. \cite{Folling2005}).  We attribute at least some of this discrepancy to collisions during the ballistic expansion of the system, which modify some atoms' trajectories, removing those atoms from the correlation features; errors in number calibration could also contribute.

The data are plotted along with a red dashed line showing the change in peak area expected if it were due solely to the finite size of the Mott domain, which we calculated in a LDA using the 2D MI phase diagram from ref.~\cite{Elstner1999}.  After we scale the model by a factor of $0.45$ to account for the overall decrease in the measured correlation-signal, it agrees to within our uncertainties.

The width of the correlation peaks are shown in Fig.~\ref{NoiseArea}B where the symbols are the measured RMS peak-widths from a Gaussian fit, and the red dashed line is the expected peak-width for a pure MI with size given by our LDA model.  At small $t/U$, the experimental data saturate to about $0.045\kr$, compared with the $0.015\kr$ expected in our model.  This saturation is due to at least two effects: (1) the finite resolution of our optical system (see Methods supporting section), and (2) the $\sim15\micron$ initial radius of the sample.  We estimate that each of these effects would separately limit the measured peak-width to about $0.03$ and $0.04\ \kr$, respectively.  (The width may also be influenced by mean field during expansion.)  The black dashed line shows the modeled peak-width added in quadrature with a width of $0.05\ \kr$, the quadrature sum of the effects described above.  The width of the correlation peak appears to increase more rapidly than expected from size arguments alone.  This suggests that the peak-width may depend on the detailed properties of the MI state (as the momentum distributions do).  It is also possible the the finite temperature of our system may influence the peak-widths, or that our masking procedure may introduce an unknown systematic error in the peak-width (in spite of evidence to the contrary: see the supporting Methods section).

In our simple calculations (valid to first order in $t/U$), the area and width of the correlation peaks depend only on the size of the Mott domain, computed in a LDA.  More sophisticated theoretical techniques can be applied to this problem, indeed explicit numerical calculations for a harmonically confined 2D Bose-Hubbard model verify that the peak-width increases as $t/U$ approaches the MI-SF critical point from the MI side~\cite{Scarola2006}.  Additionally, these authors find that when the SF phase first appears in the core of the MI (as in our experiment) the peak-width increases more rapidly as $t/U$ approaches the critical value~\cite{ScarolaPrivate}.  Numerical results for 1D systems also show a increase in the area as $t/U$ increases to the critical point~\cite{Rey2005, ReyPrivate}.  The parallels between our measurements and these numerical results indicate the need for further study of noise correlations near the MI-transition.

In this paper we have demonstrated a remarkable agreement between experiment and theory describing the momentum distribution in a 2D MI over a wide range of conditions, and to second order in perturbation theory.  Additionally, we see that correlations in the atom shot noise can be a tool for probing the SF-MI phase transition, and yield information about the fraction of the system in the MI.  Even when the momentum distribution is featureless, the noise-correlations show the lattice structure and indicate system size.   This adds support to proposals to identify the phases of extended Bose-Hubbard models (including a possible supersolid phase), using a combination of momentum and noise-correlation measurements~\cite{Scarola2006}.  

We would like to thank V. W. Scarola, E. Demler, A. M. Rey, and C. Williams for conversations, M. Anderlini and J.~Sebby-Strabley for lattice setup.  We acknowledge the financial support of ARDA, and ONR; and I.B.S. thanks the NIST/NRC fellowship program.

\bibliography{main} 

\begin{thebibliography}{27}
\expandafter\ifx\csname natexlab\endcsname\relax\def\natexlab#1{#1}\fi
\expandafter\ifx\csname bibnamefont\endcsname\relax
  \def\bibnamefont#1{#1}\fi
\expandafter\ifx\csname bibfnamefont\endcsname\relax
  \def\bibfnamefont#1{#1}\fi
\expandafter\ifx\csname citenamefont\endcsname\relax
  \def\citenamefont#1{#1}\fi
\expandafter\ifx\csname url\endcsname\relax
  \def\url#1{\texttt{#1}}\fi
\expandafter\ifx\csname urlprefix\endcsname\relax\def\urlprefix{URL }\fi
\providecommand{\bibinfo}[2]{#2}
\providecommand{\eprint}[2][]{\url{#2}}

\bibitem[{\citenamefont{Paredes et~al.}(2004)\citenamefont{Paredes, Widera,
  Murg, Mandel, F{\"o}lling, Cirac, Shlyapnikov, H{\"a}nsch, , and
  Bloch}}]{Paredes2004}
\bibinfo{author}{\bibfnamefont{B.}~\bibnamefont{Paredes}},
  \bibinfo{author}{\bibfnamefont{A.}~\bibnamefont{Widera}},
  \bibinfo{author}{\bibfnamefont{V.}~\bibnamefont{Murg}},
  \bibinfo{author}{\bibfnamefont{O.}~\bibnamefont{Mandel}},
  \bibinfo{author}{\bibfnamefont{S.}~\bibnamefont{F{\"o}lling}},
  \bibinfo{author}{\bibfnamefont{I.}~\bibnamefont{Cirac}},
  \bibinfo{author}{\bibfnamefont{G.~V.} \bibnamefont{Shlyapnikov}},
  \bibinfo{author}{\bibfnamefont{T.~W.} \bibnamefont{H{\"a}nsch}}, ,
  \bibnamefont{and} \bibinfo{author}{\bibfnamefont{I.}~\bibnamefont{Bloch}},
  \bibinfo{journal}{Nature} \textbf{\bibinfo{volume}{429}},
  \bibinfo{pages}{277} (\bibinfo{year}{2004}).

\bibitem[{\citenamefont{Kinoshita et~al.}(2004)\citenamefont{Kinoshita, Wenger,
  and Weiss}}]{Kinoshita2004}
\bibinfo{author}{\bibfnamefont{T.}~\bibnamefont{Kinoshita}},
  \bibinfo{author}{\bibfnamefont{T.~R.} \bibnamefont{Wenger}},
  \bibnamefont{and} \bibinfo{author}{\bibfnamefont{D.~S.} \bibnamefont{Weiss}},
  \bibinfo{journal}{Science} \textbf{\bibinfo{volume}{305}},
  \bibinfo{pages}{1125} (\bibinfo{year}{2004}).

\bibitem[{\citenamefont{Greiner et~al.}(2002)\citenamefont{Greiner, Mandel,
  Esslinger, H{\"a}nsch, and Bloch}}]{Greiner2002}
\bibinfo{author}{\bibfnamefont{M.}~\bibnamefont{Greiner}},
  \bibinfo{author}{\bibfnamefont{O.}~\bibnamefont{Mandel}},
  \bibinfo{author}{\bibfnamefont{T.}~\bibnamefont{Esslinger}},
  \bibinfo{author}{\bibfnamefont{T.}~\bibnamefont{H{\"a}nsch}},
  \bibnamefont{and} \bibinfo{author}{\bibfnamefont{I.}~\bibnamefont{Bloch}},
  \bibinfo{journal}{Nature} \textbf{\bibinfo{volume}{415}}, \bibinfo{pages}{39}
  (\bibinfo{year}{2002}).

\bibitem[{\citenamefont{St{\"o}ferle et~al.}(2004)\citenamefont{St{\"o}ferle,
  Moritz, Schori, K{\"o}hl, and Esslinger}}]{Stoferle2004}
\bibinfo{author}{\bibfnamefont{T.}~\bibnamefont{St{\"o}ferle}},
  \bibinfo{author}{\bibfnamefont{H.}~\bibnamefont{Moritz}},
  \bibinfo{author}{\bibfnamefont{C.}~\bibnamefont{Schori}},
  \bibinfo{author}{\bibfnamefont{M.}~\bibnamefont{K{\"o}hl}}, \bibnamefont{and}
  \bibinfo{author}{\bibfnamefont{T.}~\bibnamefont{Esslinger}},
  \bibinfo{journal}{Phys. Rev. Lett.} \textbf{\bibinfo{volume}{92}},
  \bibinfo{pages}{130403} (\bibinfo{year}{2004}).

\bibitem[{\citenamefont{K{\"o}hl et~al.}(2005)\citenamefont{K{\"o}hl, Moritz,
  St{\"o}ferle, Schori, and Esslinger}}]{Kohl2005}
\bibinfo{author}{\bibfnamefont{M.}~\bibnamefont{K{\"o}hl}},
  \bibinfo{author}{\bibfnamefont{H.}~\bibnamefont{Moritz}},
  \bibinfo{author}{\bibfnamefont{T.}~\bibnamefont{St{\"o}ferle}},
  \bibinfo{author}{\bibfnamefont{C.}~\bibnamefont{Schori}}, \bibnamefont{and}
  \bibinfo{author}{\bibfnamefont{T.}~\bibnamefont{Esslinger}},
  \bibinfo{journal}{Journal of Low Temperature Physics}
  \textbf{\bibinfo{volume}{138}}, \bibinfo{pages}{635} (\bibinfo{year}{2005}).

\bibitem[{\citenamefont{Altman et~al.}(2004)\citenamefont{Altman, Demler, and
  Lukin}}]{Altman2004}
\bibinfo{author}{\bibfnamefont{E.}~\bibnamefont{Altman}},
  \bibinfo{author}{\bibfnamefont{E.}~\bibnamefont{Demler}}, \bibnamefont{and}
  \bibinfo{author}{\bibfnamefont{M.~D.} \bibnamefont{Lukin}},
  \bibinfo{journal}{Phys. Rev. A} \textbf{\bibinfo{volume}{70}},
  \bibinfo{pages}{013603} (\bibinfo{year}{2004}).

\bibitem[{\citenamefont{F{\"o}lling et~al.}(2005)\citenamefont{F{\"o}lling,
  Gerbier, Widera, Mandel, Gericke, and Bloch}}]{Folling2005}
\bibinfo{author}{\bibfnamefont{S.}~\bibnamefont{F{\"o}lling}},
  \bibinfo{author}{\bibfnamefont{F.}~\bibnamefont{Gerbier}},
  \bibinfo{author}{\bibfnamefont{A.}~\bibnamefont{Widera}},
  \bibinfo{author}{\bibfnamefont{O.}~\bibnamefont{Mandel}},
  \bibinfo{author}{\bibfnamefont{T.}~\bibnamefont{Gericke}}, \bibnamefont{and}
  \bibinfo{author}{\bibfnamefont{I.}~\bibnamefont{Bloch}},
  \bibinfo{journal}{Nature} \textbf{\bibinfo{volume}{434}},
  \bibinfo{pages}{481} (\bibinfo{year}{2005}).

\bibitem[{\citenamefont{Greiner et~al.}(2005)\citenamefont{Greiner, Regal,
  Stewart, and Jin}}]{Greiner2005}
\bibinfo{author}{\bibfnamefont{M.}~\bibnamefont{Greiner}},
  \bibinfo{author}{\bibfnamefont{C.~A.} \bibnamefont{Regal}},
  \bibinfo{author}{\bibfnamefont{J.~T.} \bibnamefont{Stewart}},
  \bibnamefont{and} \bibinfo{author}{\bibfnamefont{D.~S.} \bibnamefont{Jin}},
  \bibinfo{journal}{Phys. Rev. Lett.} \textbf{\bibinfo{volume}{94}},
  \bibinfo{pages}{110401} (\bibinfo{year}{2005}).

\bibitem[{\citenamefont{Schellekens et~al.}(2005)\citenamefont{Schellekens,
  Hoppeler, Perrin, Gomes, Boiron, Aspect, and Westbrook}}]{Schellekens2005}
\bibinfo{author}{\bibfnamefont{M.}~\bibnamefont{Schellekens}},
  \bibinfo{author}{\bibfnamefont{R.}~\bibnamefont{Hoppeler}},
  \bibinfo{author}{\bibfnamefont{A.}~\bibnamefont{Perrin}},
  \bibinfo{author}{\bibfnamefont{J.~V.} \bibnamefont{Gomes}},
  \bibinfo{author}{\bibfnamefont{D.}~\bibnamefont{Boiron}},
  \bibinfo{author}{\bibfnamefont{A.}~\bibnamefont{Aspect}}, \bibnamefont{and}
  \bibinfo{author}{\bibfnamefont{C.~I.} \bibnamefont{Westbrook}},
  \bibinfo{journal}{Science} \textbf{\bibinfo{volume}{310}},
  \bibinfo{pages}{648} (\bibinfo{year}{2005}).

\bibitem[{\citenamefont{Jaksch et~al.}(1998)\citenamefont{Jaksch, Bruder,
  Cirac, Gardiner, and Zoller}}]{Jaksch1998}
\bibinfo{author}{\bibfnamefont{D.}~\bibnamefont{Jaksch}},
  \bibinfo{author}{\bibfnamefont{C.}~\bibnamefont{Bruder}},
  \bibinfo{author}{\bibfnamefont{J.~I.} \bibnamefont{Cirac}},
  \bibinfo{author}{\bibfnamefont{C.~W.} \bibnamefont{Gardiner}},
  \bibnamefont{and} \bibinfo{author}{\bibfnamefont{P.}~\bibnamefont{Zoller}},
  \bibinfo{journal}{Physical Review Letters} \textbf{\bibinfo{volume}{81}},
  \bibinfo{pages}{3108} (\bibinfo{year}{1998}).

\bibitem[{\citenamefont{Batrouni et~al.}(2002)\citenamefont{Batrouni, Rousseau,
  Scalettar, Rigol, Muramatsu, Denteneer, and Troyer}}]{Batrouni2002}
\bibinfo{author}{\bibfnamefont{G.~G.} \bibnamefont{Batrouni}},
  \bibinfo{author}{\bibfnamefont{V.}~\bibnamefont{Rousseau}},
  \bibinfo{author}{\bibfnamefont{R.~T.} \bibnamefont{Scalettar}},
  \bibinfo{author}{\bibfnamefont{M.}~\bibnamefont{Rigol}},
  \bibinfo{author}{\bibfnamefont{A.}~\bibnamefont{Muramatsu}},
  \bibinfo{author}{\bibfnamefont{P.~J.~H.} \bibnamefont{Denteneer}},
  \bibnamefont{and} \bibinfo{author}{\bibfnamefont{M.}~\bibnamefont{Troyer}},
  \bibinfo{journal}{Phys. Rev. Lett.} \textbf{\bibinfo{volume}{89}}
  (\bibinfo{year}{2002}).

\bibitem[{\citenamefont{F{\"o}lling et~al.}(2006)\citenamefont{F{\"o}lling,
  Widera, M{\"u}ller, Gerbier, and Bloch}}]{Folling2006}
\bibinfo{author}{\bibfnamefont{S.}~\bibnamefont{F{\"o}lling}},
  \bibinfo{author}{\bibfnamefont{A.}~\bibnamefont{Widera}},
  \bibinfo{author}{\bibfnamefont{T.}~\bibnamefont{M{\"u}ller}},
  \bibinfo{author}{\bibfnamefont{F.}~\bibnamefont{Gerbier}}, \bibnamefont{and}
  \bibinfo{author}{\bibfnamefont{I.}~\bibnamefont{Bloch}},
  \bibinfo{journal}{arxiv:cond-mat} p. \bibinfo{pages}{0606592}
  (\bibinfo{year}{2006}).

\bibitem[{\citenamefont{Campbell et~al.}(2006)\citenamefont{Campbell, Mun,
  Boyd, Medley, Leanhardt, Marcassa, Pritchard, and Ketterle}}]{Campbell2006}
\bibinfo{author}{\bibfnamefont{G.~K.} \bibnamefont{Campbell}},
  \bibinfo{author}{\bibfnamefont{J.}~\bibnamefont{Mun}},
  \bibinfo{author}{\bibfnamefont{M.}~\bibnamefont{Boyd}},
  \bibinfo{author}{\bibfnamefont{P.}~\bibnamefont{Medley}},
  \bibinfo{author}{\bibfnamefont{A.~E.} \bibnamefont{Leanhardt}},
  \bibinfo{author}{\bibfnamefont{L.}~\bibnamefont{Marcassa}},
  \bibinfo{author}{\bibfnamefont{D.~E.} \bibnamefont{Pritchard}},
  \bibnamefont{and} \bibinfo{author}{\bibfnamefont{W.}~\bibnamefont{Ketterle}},
  \bibinfo{journal}{arXiv:cond-mat} \textbf{\bibinfo{volume}{0606642}}
  (\bibinfo{year}{2006}).

\bibitem[{\citenamefont{Fisher et~al.}(1989)\citenamefont{Fisher, Weichman,
  Grinstein, and Fisher}}]{Fisher1989}
\bibinfo{author}{\bibfnamefont{M.~P.~A.} \bibnamefont{Fisher}},
  \bibinfo{author}{\bibfnamefont{P.~B.} \bibnamefont{Weichman}},
  \bibinfo{author}{\bibfnamefont{G.}~\bibnamefont{Grinstein}},
  \bibnamefont{and} \bibinfo{author}{\bibfnamefont{D.~S.}
  \bibnamefont{Fisher}}, \bibinfo{journal}{Physical Review B}
  \textbf{\bibinfo{volume}{40}}, \bibinfo{pages}{546} (\bibinfo{year}{1989}).

\bibitem[{\citenamefont{Elstner and Monien}(1999)}]{Elstner1999}
\bibinfo{author}{\bibfnamefont{N.}~\bibnamefont{Elstner}} \bibnamefont{and}
  \bibinfo{author}{\bibfnamefont{H.}~\bibnamefont{Monien}},
  \bibinfo{journal}{arXiv:cond-mat} p. \bibinfo{pages}{9905367}
  (\bibinfo{year}{1999}).

\bibitem[{\citenamefont{Wessel et~al.}(2004)\citenamefont{Wessel, Alet, Troyer,
  and Batrouni}}]{Wessel2004}
\bibinfo{author}{\bibfnamefont{S.}~\bibnamefont{Wessel}},
  \bibinfo{author}{\bibfnamefont{F.}~\bibnamefont{Alet}},
  \bibinfo{author}{\bibfnamefont{M.}~\bibnamefont{Troyer}}, \bibnamefont{and}
  \bibinfo{author}{\bibfnamefont{G.~G.} \bibnamefont{Batrouni}},
  \bibinfo{journal}{Physical Review A} \textbf{\bibinfo{volume}{70}},
  \bibinfo{pages}{053615} (\bibinfo{year}{2004}).

\bibitem[{\citenamefont{Gerbier
  et~al.}(2005{\natexlab{a}})\citenamefont{Gerbier, Widera, F{\"o}lling,
  Mandel, Gericke, and Bloch}}]{Gerbier2005}
\bibinfo{author}{\bibfnamefont{F.}~\bibnamefont{Gerbier}},
  \bibinfo{author}{\bibfnamefont{A.}~\bibnamefont{Widera}},
  \bibinfo{author}{\bibfnamefont{S.}~\bibnamefont{F{\"o}lling}},
  \bibinfo{author}{\bibfnamefont{O.}~\bibnamefont{Mandel}},
  \bibinfo{author}{\bibfnamefont{T.}~\bibnamefont{Gericke}}, \bibnamefont{and}
  \bibinfo{author}{\bibfnamefont{I.}~\bibnamefont{Bloch}},
  \bibinfo{journal}{Phys. Rev. Lett.} \textbf{\bibinfo{volume}{95}},
  \bibinfo{pages}{050404} (\bibinfo{year}{2005}{\natexlab{a}}).

\bibitem[{\citenamefont{Gerbier
  et~al.}(2005{\natexlab{b}})\citenamefont{Gerbier, Widera, F{\"o}lling,
  Mandel, Gericke, and Bloch}}]{Gerbier2005a}
\bibinfo{author}{\bibfnamefont{F.}~\bibnamefont{Gerbier}},
  \bibinfo{author}{\bibfnamefont{A.}~\bibnamefont{Widera}},
  \bibinfo{author}{\bibfnamefont{S.}~\bibnamefont{F{\"o}lling}},
  \bibinfo{author}{\bibfnamefont{O.}~\bibnamefont{Mandel}},
  \bibinfo{author}{\bibfnamefont{T.}~\bibnamefont{Gericke}}, \bibnamefont{and}
  \bibinfo{author}{\bibfnamefont{I.}~\bibnamefont{Bloch}},
  \bibinfo{journal}{Physical Review A} \textbf{\bibinfo{volume}{72}},
  \bibinfo{pages}{053606} (\bibinfo{year}{2005}{\natexlab{b}}).

\bibitem[{\citenamefont{Spielman et~al.}(2006)\citenamefont{Spielman, Johnson,
  Huckans, Fertig, Rolston, Phillips, and Porto}}]{Spielman2006}
\bibinfo{author}{\bibfnamefont{I.~B.} \bibnamefont{Spielman}},
  \bibinfo{author}{\bibfnamefont{P.~R.} \bibnamefont{Johnson}},
  \bibinfo{author}{\bibfnamefont{J.~H.} \bibnamefont{Huckans}},
  \bibinfo{author}{\bibfnamefont{C.~D.} \bibnamefont{Fertig}},
  \bibinfo{author}{\bibfnamefont{S.~L.} \bibnamefont{Rolston}},
  \bibinfo{author}{\bibfnamefont{W.~D.} \bibnamefont{Phillips}},
  \bibnamefont{and} \bibinfo{author}{\bibfnamefont{J.~V.} \bibnamefont{Porto}},
  \bibinfo{journal}{Phys. Rev. A} \textbf{\bibinfo{volume}{73}},
  \bibinfo{pages}{020702(R)} (\bibinfo{year}{2006}).

\bibitem[{\citenamefont{Sebby-Strabley
  et~al.}(2006)\citenamefont{Sebby-Strabley, Anderlini, Jessen, and
  Porto}}]{Sebby-Strabley2006}
\bibinfo{author}{\bibfnamefont{J.}~\bibnamefont{Sebby-Strabley}},
  \bibinfo{author}{\bibfnamefont{M.}~\bibnamefont{Anderlini}},
  \bibinfo{author}{\bibfnamefont{P.~S.} \bibnamefont{Jessen}},
  \bibnamefont{and} \bibinfo{author}{\bibfnamefont{J.~V.} \bibnamefont{Porto}},
  \bibinfo{journal}{Phys. Rev. A} \textbf{\bibinfo{volume}{73}},
  \bibinfo{pages}{033605} (\bibinfo{year}{2006}).

\bibitem[{\citenamefont{Ovchinnikov et~al.}(1998)\citenamefont{Ovchinnikov,
  M\"uller, Doery, Vredenbregt, Helmerson, Rolston, and
  Phillips}}]{Ovchinnikov1998}
\bibinfo{author}{\bibfnamefont{Y.~B.} \bibnamefont{Ovchinnikov}},
  \bibinfo{author}{\bibfnamefont{J.~H.} \bibnamefont{M\"uller}},
  \bibinfo{author}{\bibfnamefont{M.~R.} \bibnamefont{Doery}},
  \bibinfo{author}{\bibfnamefont{E.~J.~D.} \bibnamefont{Vredenbregt}},
  \bibinfo{author}{\bibfnamefont{K.}~\bibnamefont{Helmerson}},
  \bibinfo{author}{\bibfnamefont{S.~L.} \bibnamefont{Rolston}},
  \bibnamefont{and} \bibinfo{author}{\bibfnamefont{W.~D.}
  \bibnamefont{Phillips}}, \bibinfo{journal}{Phys. Rev. Lett.}
  \textbf{\bibinfo{volume}{83}}, \bibinfo{pages}{284} (\bibinfo{year}{1998}).

\bibitem[{\citenamefont{Sengupta and Dupuis}(2005)}]{Sengupta2005}
\bibinfo{author}{\bibfnamefont{K.}~\bibnamefont{Sengupta}} \bibnamefont{and}
  \bibinfo{author}{\bibfnamefont{N.}~\bibnamefont{Dupuis}},
  \bibinfo{journal}{Physical Review A} \textbf{\bibinfo{volume}{71}},
  \bibinfo{pages}{033629} (\bibinfo{year}{2005}).

\bibitem[{\citenamefont{Rey et~al.}(2005)\citenamefont{Rey, Satija, and
  Clark}}]{Rey2005}
\bibinfo{author}{\bibfnamefont{A.~M.} \bibnamefont{Rey}},
  \bibinfo{author}{\bibfnamefont{I.~I.} \bibnamefont{Satija}},
  \bibnamefont{and} \bibinfo{author}{\bibfnamefont{C.~W.} \bibnamefont{Clark}}
  (\bibinfo{year}{2005}), \bibinfo{note}{cond-mat/0511700}.

\bibitem[{\citenamefont{Scarola et~al.}(2006)\citenamefont{Scarola, Demler, and
  Sarma}}]{Scarola2006}
\bibinfo{author}{\bibfnamefont{V.~W.} \bibnamefont{Scarola}},
  \bibinfo{author}{\bibfnamefont{E.}~\bibnamefont{Demler}}, \bibnamefont{and}
  \bibinfo{author}{\bibfnamefont{S.~D.} \bibnamefont{Sarma}},
  \bibinfo{journal}{Physical Review A} \textbf{\bibinfo{volume}{73}},
  \bibinfo{pages}{051601(R)} (\bibinfo{year}{2006}).

\bibitem[{\citenamefont{Scarola}()}]{ScarolaPrivate}
\bibinfo{author}{\bibfnamefont{V.~W.} \bibnamefont{Scarola}},
  \bibinfo{howpublished}{Private communication}.

\bibitem[{\citenamefont{Rey}()}]{ReyPrivate}
\bibinfo{author}{\bibfnamefont{A.~M.} \bibnamefont{Rey}},
  \bibinfo{howpublished}{Private communication}.

\bibitem[{\citenamefont{Gericke et~al.}(2006)\citenamefont{Gericke, Gerbier,
  Widera, F{\"o}lling, Mandel, and Bloch}}]{Gericke2006}
\bibinfo{author}{\bibfnamefont{T.}~\bibnamefont{Gericke}},
  \bibinfo{author}{\bibfnamefont{F.}~\bibnamefont{Gerbier}},
  \bibinfo{author}{\bibfnamefont{A.}~\bibnamefont{Widera}},
  \bibinfo{author}{\bibfnamefont{S.}~\bibnamefont{F{\"o}lling}},
  \bibinfo{author}{\bibfnamefont{O.}~\bibnamefont{Mandel}}, \bibnamefont{and}
  \bibinfo{author}{\bibfnamefont{I.}~\bibnamefont{Bloch}}
  (\bibinfo{year}{2006}), \bibinfo{note}{cond-mat/0603590}.

\end{thebibliography}

\pagebreak

\section{Methods}

\subsection{Image acquisition}

Once the confining potentials are removed the atom cloud expands for a  TOF ($20.1\ms$ for the data probing for noise-correlations, and $29.1\ms$ otherwise).  The atoms are then illuminated for $t_{\rm pulse} = 50 \us$ by a circularly polarized probe beam resonant with the $\left|F=2,m_F=2\right>\rightarrow\left|F=3,m_F=3\right>$ cycling transition of the $\Rb87$ D2 line.  The probe beam was carefully aligned parallel with the vertical lattice, and approximately with a $\approx 0.2\ {\rm mT}$ bias field, with intensity $I=0.2 I_0$, where $I_0$ is the resonant saturation intensity.  During the imaging pulse an atom with lifetime $\Gamma$ will scatter $N_\gamma=\Gamma t_{\rm pulse}/ [2 (1+I_0/I)] \approx 160$ photons out of the beam.  We image the shadow cast by the atoms onto a CCD array.  

Our ultimate imaging resolution is set by several effects: the resolution limit of the optical system (diffraction and aberration), the finite size of the CCD pixels, and the movement of the atoms during imaging (which is negligible in our case).   

We measure a RMS optical resolution of $5\micron$ by fitting a Gaussian to the spatial power spectra of atom shot-noise.  The diffraction limit of our optical system is predominately determined by a pair of lenses with $f/\#=2.5$.  At $780\nm$ this pair gives a diffraction limited spot with a $2.5\micron$ RMS radius.  After the $6\times$ magnification of our imaging system, the CCD has an effective pixel size of $2.5\micron$ at the cloud.  Based on the quadrature sum of these terms, a single atom should be imaged within a circle with a RMS radius of about $4\micron$.  

While the discussion in the body of the text focused upon fundamental atom shot noise, any real experiment is also beset with technical noise.  In our absorption images, the CCD is illuminated by a probe beam that is only slightly attenuated by the atoms.  The CCD array counts the remaining photons, which arrive stochastically -- photon shot noise -- which acts as technical noise.  Second, a readout error is introduced when the photo-electrons in each CCD bin are counted.  Together these give an expected RMS background statistical uncertainty in our reported optical depths of $0.02$.  In our data, a resolution-limited spot might contain $\approx 25$ atoms, which has a $5$ atom counting uncertainty giving rise to an optical depth change of $0.02$.  Only a fraction of that noise is correlated, requiring us to average many images to extract the correlated noise signal.

\subsection{Image processing}

The atom noise signal represents the measured shot to shot fluctuations of the momentum distribution around the average distribution. 

The noise-correlations in the momentum distribution are a subtle feature which can easily be mimicked by correlations in the average momentum distribution if the appropriate average is not subtracted.  Therefore, it is critical to accurately determine an average distribution, $\nbarexpt$, to subtract from the measured distributions.  Ideally, $\nbarexpt$ is determined by averaging a large number of measured distributions under identical conditions.  In practice, fluctuations in initial conditions (lattice depth: $\pm 10\%$, atom number: $\pm20\%$, transverse velocity: $\pm 250\micron/{\rm s}$, and etc.) make it difficult to precisely define $\nbarexpt$.  As a result, spurious ``ghosts'' of diffractive structure are replicated in the correlation-data due to imperfect removal of the atom background.  This problem is not evident except when sharp diffractive peaks appear in the momentum distribution, i.e., $t/U\gtrsim \ToverUC$.

To mitigate this, each image is processed as follows: we first fit to an empirical multi-parameter function which includes diffraction peaks.  When subtracted from the initial image, this removes the average atom density (ideally leaving only the noise signal including both atom shot-noise and technical noise).  However, because our fitting function is imperfect, some fine structure still persists at the locations of the diffraction peaks after subtraction.  If ignored, these features give rise to a spurious signal in the ACF exactly at the wave-vectors of interest.  We note that these features exist only at the location of diffraction peaks and are separated by multiples of $2\kr$.

We remove the effect of this localized structure on the ACF by masking circular regions with a $35\micron$ radius radius centered on the location of the diffraction peaks (the diffraction peaks have a typical radius ranging from $15$ to $20\micron$).  We find both empirically (when there is no diffractive structure, i.e., at very small $t/U$), and in simulation that such masks have no effect on the normalized ACF. (F\"olling {\it et al} used a similar technique, but only deep in the MI phase, to show that the correlation-peaks do not result from spurious effects of remnant diffraction~\cite{Folling2005}.)  We verified that the size of the mask had little effect on the final ACF when the radius ranged from $0$ to $45\micron$ when $t/U$ is small (the least diffractive data) and from $30$ to $45\micron$ when $t/U$ is large (the most diffractive data).  For masks larger than about $45\micron$, we could no longer distinguish the desired correlation features from noise in the most diffractive data.

When left unmasked, the spurious diffractive signal introduces excess correlations which (1) increase the overall magnitude of the correlation peaks, and (2) increase the width of the correlation peaks.

There are two additional imaging artifacts which we remove.  First we 
remove a high spatial frequency component due to noise in the readout electronics, which is present in every image (with variable phase and amplitude).  Second, we remove a low frequency  modulation ($\approx 3$ oscillations over the whole image) due to diffraction fringes of the probe beam.

\end{document}